\documentclass[12pt]{iopart}

\usepackage{amssymb}
\usepackage{amsbsy}
\usepackage{graphicx}
\usepackage{float}
\usepackage{color}

\newcommand{\bq}{\begin{eqnarray}}
\newcommand{\eq}{\end{eqnarray}}
\newcommand{\rr}{\mathbf{r}}
\newcommand{\qq}{\mathbf{q}}

\begin{document}
\title{Two Component Plasma in a Flamm's Paraboloid}  

\author{Riccardo Fantoni}
\ead{rfantoni27@sun.ac.za}
\address{National Institute for Theoretical Physics (NITheP) and
Institute of Theoretical Physics,  University of
Stellenbosch, Stellenbosch 7600, South Africa} 


\date{\today}

\begin{abstract}
The two component plasma (TCP) living in a Flamm's paraboloid is
studied at a value of the coupling constant $\Gamma=2$ for which an
analytic expression for the grand canonical partition function is
available. Two cases are considered, the plasma in the half surface with
an insulating horizon and the plasma in the whole surface. The
Green's function equation necessary to determine 
the $n$-particle truncated correlation functions is explicitly
found. In both cases this proves too complicated to be solved 
analytically. So we present the method of solution
reducing the problem to finding the two linearly independent solutions
of a linear homogeneous second order ordinary differential equation
with polynomial coefficients of high degrees. In the flat limit one recovers
the solution for the structure of the TCP in a plane in the first case
but the collapse of opposite charges at the horizon makes the
structure of the plasma physically not well defined in the second
case.  
\end{abstract}


\maketitle
\section{Introduction}
\label{sec:introduction}

The two component plasma (TCP) is a neutral mixture of point wise particles
of charge $\pm q$. The equation of state of the TCP
living in a plane is known since the work of Salzberg and Prager
\cite{Salzberg63}. In the plasma the attraction between oppositely
charged particles competes with the thermal motion and makes the
partition function of the finite system diverge when 
$\Gamma = \beta q^2 \ge 2$, where $\beta=1/k_BT$ with $k_B$ Boltzmann
constant. The system becomes unstable against the collapse of pairs
of oppositely charged particles, and as a consequence all
thermodynamic quantities diverge, so that the point particle model is
well behaved only for $\Gamma < 2$ \cite{Hauge71} when the Boltzmann
factor for unlike particles is integrable at small separations of the
charges. In this case rescaling the particles coordinates so as 
to stay in the unit disk one easily proves that the grand canonical
partition function is a function of
$\sqrt{\lambda_-\lambda_+}V^{(1-\Gamma/4)}$, where $V$ is the volume
of the plasma and $\lambda_\pm$ the fugacities of 
the two charge species, and as a consequence the equation of state is  
$\beta P=n(1-\Gamma/4)$ where $n=n_++n_-$ is the total particle number
density. It also follows that the ratio
$\sqrt{\lambda_-\lambda_+}/n^{(1-\Gamma/4)}$ must be a function of
$\Gamma$ only in the thermodynamic limit \cite{Samaj2000}. 
However, if the collapse is avoided by some short range
repulsion (hard cores for instance), the model remains well defined
for lower temperatures. Then, for $\Gamma > 4$ 
the long range Coulomb attraction binds positive and negative
particles in pairs of finite polarizability. Thus, at some critical
value $\Gamma_c \sim 4$ the system undergoes the Kosterlitz-Thouless
transition \cite{Kosterlitz1973} between a high temperature ($\Gamma <
4$) conductive phase 
and a low temperature ($\Gamma > 4$) dielectric phase. The same
behavior also occurs in the TCP living in one 
dimension \cite{Lenard1961}. 

The structure of the TCP living in a plane at the
special value $\Gamma=2$ of the coupling constant is solvable exactly
analytically \cite{Cornu87}. Through the use of an external potential it
has also been studied in various confined geometries
\cite{Cornu89,Forrester91,Tellez02,Merchan04} and in a gravitational
field \cite{Tellez97,Tellez98}.  

On curved surfaces it has only been studied in surfaces of constant
curvature as the sphere \cite{Forrester1992,Forrester1996} and the
pseudosphere \cite{Jancovici1998}. Unlike the one component plasma
where the properties of the Vandermonde determinant allowed the
analytical solution a Cauchy identity is used for the solution of the
TCP. Unlike in the one component case where the
solution was possible for the plasma confined in a region of the
surface now this is not possible, anymore, without the use of an
external potential. In these cases the external potential is rather given by
$-(\Gamma/q^2)\ln\sqrt g$ where $g$ is the determinant of the metric
tensor. 

In this work we study the TCP at the special value
$\Gamma=2$ of the coupling constant in a Flamm's paraboloid, a surface
of non constant curvature. In this surface the one component plasma
\cite{Fantoni2008} has been studied in various situations: confined to
one ``universe'' by an insulating or a grounded perfect conductor
``horizon'' or in the whole surface. Here we will restrict ourselves
only to the first and last situation.

On a curved surface, even though the finite system partition function
will still be finite for $\Gamma<2$ since the surface is locally flat,
the structure will change respect to the flat case. The purpose of the
present work is to see how it changes for the special case $\Gamma=2$
where an exact analytical solution is possible.

\section{The Flamm's paraboloid $\cal S$}
\label{sec:surface}
In this work, we want to study a two dimensional TCP
in the Flamm's paraboloid. This is a Riemannian surface ${\cal
  S}$ with the following metric 
\begin{eqnarray}  \label{metric} 
d\mathbf{s}^2=g_{\mu\nu}dx^\mu dx^\nu=
\left(1-\frac{2M}{r}\right)^{-1}dr^2+r^2d\varphi^2~.
\end{eqnarray}
or $g_{rr}=1/(1-2M/r),g_{\varphi\varphi}=r^2$, and $g_{r\varphi}=0$.

The Flamm's paraboloid is an embeddable surface in the
three-dimensional Euclidean space. It is composed by two identical
``universes'': ${\cal S}_+$ and ${\cal S}_-$. These are both multiply
connected surfaces connected by a hole of radius $2M$. We will from
now on call the $r=2M$ region of the surface its ``horizon''. 
The scalar curvature is $R= -2M/r^3$.

The system of coordinates $(r,\varphi)$ with the metric~(\ref{metric})
has the disadvantage that it requires two charts to cover the whole
surface ${\cal S}$. It can be more convenient to use the variable
\begin{eqnarray} \label{u}
u=\pm\sqrt{\frac{r}{2M}-1}
\end{eqnarray}
instead of $r$. This gives the following metric when using the system
of coordinates $(u,\varphi)$,
\begin{eqnarray}
d\mathbf{s}^2=4M^2(1+u^2)\left[4\,du^2+(1+u^2)\,d\varphi^2\right]~.
\end{eqnarray}
The region $u>0$ corresponds to ${\cal S}_+$ and the region $u<0$
to ${\cal S}_{-}$.

The Laplace-Beltrami operator is
\begin{eqnarray} \nonumber
\Delta f&=&
\frac{1}{\sqrt{g}}\frac{\partial}{\partial q^\mu}
\left(\sqrt{g}\,g^{\mu\nu}\frac{\partial}{\partial q^\nu}\right)f\\
&=&\left[\left(1-\frac{2M}{r}\right)\partial^2_r
+\frac{1}{r^2}\partial^2_\varphi
+\left(\frac{1}{r}-\frac{M}{r^2}\right)\partial_r\right]
f~,
\end{eqnarray}
where $\qq\equiv(r,\varphi)$. Finding the Green's function of the
Laplacian, naturally \cite{Fantoni2008} leads to consider the system
of coordinates $(s,\varphi)$, with 
\begin{equation}
  s=(\sqrt{u^2+1}+u)^{2}
  \,.
\end{equation}
The range for the variable $s$ is $\left]0,+\infty\right[$. The lower
paraboloid ${\cal S}_{-}$ corresponds to the region $0<s<1$ and
the upper one ${\cal S}_{+}$ to the region $s>1$. A point in the
upper paraboloid with coordinate $(s,\varphi)$ has a mirror image
by reflection ($u\to -u$) in the lower paraboloid, with
coordinates $(1/s,\varphi)$, since if
\begin{equation}
  s=(\sqrt{u^2+1}+u)^{2}
\end{equation}
then
\begin{equation}
  \frac{1}{s}=(\sqrt{u^2+1}-u)^{2}
  \,.
\end{equation}
In the upper paraboloid ${\cal S}_{+}$, the new coordinate $s$ can be
expressed in terms of the original one, $r$, as
\begin{equation}
s=\frac{(\sqrt{r}+\sqrt{r-2M})^2}{2M}  
\,.
\end{equation}

Using this system of coordinates, the metric takes the form of a
flat metric multiplied by a conformal factor
\begin{equation}
  \label{eq:metric-in-x}
  d\mathbf{s}^2=
  \frac{M^2}{4}\left(1+\frac{1}{s}\right)^4
  \left(ds^2+s^2\,d\varphi^2\right)\,.
\end{equation}
The Laplacian also takes a simple form
\begin{equation}
  \Delta f =\frac{s}{\sqrt g}
  \,\Delta_{\mathrm{flat}}f
\end{equation}
where 
\begin{equation}
  \Delta_{\mathrm{flat}}f=
  \partial^2_s f
  +\frac{1}{s}\partial_s f
  +\frac{1}{s^2}\partial^2_\varphi f
\end{equation}
is the Laplacian of the flat Euclidean space $\mathbb{R}^2$.  The
square root of the determinant of the metric is now given by
$\sqrt{g}=(M/2)^2 s (1+s^{-1})^4$. 

It is useful to keep in mind the following small $M$ behaviors
\bq
s &=& 2\frac{r}{M}-2-\frac{1}{2}\frac{M}{r}-
\frac{1}{2}\left(\frac{M}{r}\right)^2-
\frac{5}{8}\left(\frac{M}{r}\right)^3+
O(M^4)~,\\
\sqrt g &=& \frac{rM}{2}+\frac{M^2}{2}+\frac{5M^3}{8r}+O(M^4)~.
\eq
\section{Coulomb potential created by a point charge}
\label{sec:coulomb}

We here summarize the results found in In Ref. \cite{Fantoni2008} on
the determination of the Coulomb potential between two charges living
in half Flamm's paraboloid with an insulating horizon and between two
charges living in the full Flamm's paraboloid.

The Coulomb potential $G(s,\varphi;s_0,\varphi_0)$ created at
$(s,\varphi)$ by a unit charge at $(s_0,\varphi_0)$ is given by
the Green's function of the Laplacian
\begin{equation}
  \label{eq:LaplaceGreen}
  \Delta G(s,\varphi;s_0,\varphi_0)
  =-2\pi \delta^{(2)}(s,\varphi;s_0,\varphi_0)
\end{equation}
with appropriate boundary conditions. The Dirac distribution is given
by
\begin{equation}
  \delta^{(2)}(s,\varphi;s_0,\varphi_0)
  =\frac{1}{\sqrt{g}}\delta(s-s_0)\delta(\varphi-\varphi_0)
\end{equation}

Notice that using the system of coordinates $(s,\varphi)$ the
Laplacian Green's function equation takes the simple form
\begin{equation}
  \label{eq:GreenLaplace-flat}
  \Delta_{\mathrm{flat}}  G(s,\varphi;s_0,\varphi_0)
    =-2\pi\frac{1}{s}\,\delta(s-s_0)\delta(\varphi-\varphi_0)
\end{equation}
which is formally the same Laplacian Green's function equation for a
flat space.

The Laplacian Green's function equation~(\ref{eq:LaplaceGreen}) can be
solved, as usual, by using the decomposition as a Fourier
series. Since Eq.~(\ref{eq:LaplaceGreen}) reduces to the flat
Laplacian Green's function equation~(\ref{eq:GreenLaplace-flat}), the
solution is the standard one
\begin{equation}
  \label{eq:Fourier}
  G(s,\varphi;s_0,\varphi_0)=
  \sum_{n=1}^{\infty}
  \frac{1}{n}\left(\frac{s_{<}}{s_{>}}\right)^{n}
  \cos\left[ n(\varphi-\varphi_0)\right]
  +g_0(s,s_0)
\end{equation}
where $s_{>}=\max(s,s_0)$ and $s_{<}=\min(s,s_0)$. 

We consider now the case of particles restricted to live in
the half surface (hs) ${\cal S}_{+}$, $s>1$, by a
hard wall located at the ``horizon'', $s=1$. The region ${\cal
  S}_{-}$, $s<1$, is empty and has the same dielectric constant as the
upper region. We want to consider a model where the interaction
potential reduces to the flat Coulomb potential in the flat limit
$M\to 0$. The solution of the Laplacian Green's function equation is
given in Fourier series by Eq.~(\ref{eq:Fourier}). The 0th order
Fourier component $g_0$ can be determined by the requirement that, in
the limit $M\to 0$, the solution reduces to the flat Coulomb potential
\begin{equation}
  G^{\mathrm{flat}}(\rr,\rr')=-\ln\frac{|\rr-\rr'|}{L} 
\end{equation}
where $L$ is an arbitrary constant length. Recalling that $s\sim 2r/M$,
when $M\to 0$, we find
\begin{equation}
  \label{eq:g0-hs}
  g_0(s,s_0)=-\ln s_{>}-\ln\frac{M}{2L}
\end{equation}
and in the half surface 
\begin{equation}
  \label{cgreen}
  G^{\mathrm{hs}}
  (s,\varphi;s_0,\varphi_0)=-\ln |z-z_0|-\ln\frac{M}{2L}~,
\end{equation}
where we defined $z=se^{i\varphi}$ and $z_0=se^{i\varphi_0}$. And
$\lim_{M\to 0}G^{\mathrm{hs}}=G^{\mathrm{flat}}$ as desired.

A similar procedure \cite{Fantoni2008} gives the Coulomb potential
between charges living in the whole surface (ws) $0<s<\infty$,  
\bq \label{vcoulombws}
G^{\mathrm{ws}}(s,\varphi;s_0,\varphi_0)=-\ln\frac{|z-z_0|}
{\sqrt{|zz_0|}}-\ln \frac{L_0}{L}~, 
\eq
with $L_0$ another length scale.
\section{The two component plasma model}
\label{sec:model}
The TCP is a neutral mixture of two species of point
charges of charge $\pm q$. In this work we want to study the plasma on
the Flamm's paraboloid. In Ref. \cite{Fantoni2008} the Coulomb
potential between two charges living in the whole paraboloid, 
on half surface with an insulating horizon, and in half surface with a
grounded horizon was found. The first and last case have been
summarized in the previous section as they will be the two cases
considered next. 

\section{TCP in the half surface with an insulating horizon}
\label{sec:tcp-hs}
When the TCP lives in the half surface with an insulating horizon the
Coulomb potential is given by Eq. (\ref{cgreen}). We will use
$u_i=s_ie^{i\varphi_i}$ and $v_j=s_je^{i\varphi_j}$ to denote the
complex coordinates of the positively and negatively charged particles
respectively, where $s=(\sqrt{r}+\sqrt{r-2M})^2/2M > 1$. Then,
following Ref. \cite{Cornu87}, we use a Cauchy identity \cite{Aitken}
to rewrite $e^{-\beta U_{2N}}$, where $U_{2N}$ is the potential energy
of a neutral system of $N$ positive and $N$ negative charges, as
follows     
\bq \nonumber
&&\exp\left\{2\sum_{i<j}\left[\ln\left|\frac{u_i-u_j}{2L/M}\right|+
\ln\left|\frac{v_i-v_j}{2L/M}\right|\right]-2\sum_{i,j}
\ln\left|\frac{u_i-v_j}{2L/M}\right|\right\}\\ \nonumber
&&=\left(\frac{2L}{M}\right)^{2N}
\left|\frac{\prod_{i<j}(u_i-u_j)(v_i-v_j)}
{\prod_{i,j}(u_i-v_j)}\right|^2\\ 
&&=\left(\frac{2L}{M}\right)^{2N}
\left|\det\left(\frac{1}{u_i-v_j}
\right)_{(i,j)\in\{1,\ldots,N\}^2}\right|^2~.
\eq
where we had to choose the particular case of a coupling constant
$\Gamma=\beta q^2=2$.

Following Ref. \cite{Cornu87}, to avoid divergences we start
from a discretized model in which two 
sub-lattices $U$ and $V$ are introduced . The positive (negative)
particles sit on the sub-lattice $U (V)$. Each lattice site is occupied
by no or one particle. 
The grand canonical partition function defined as a sum including only
neutral systems is then
\bq \label{pf0}
\Xi&=&1+\sum_{N=1}^\infty \left(\frac{2L}{M}\right)^{2N}\lambda_-^N
\lambda_+^N
\mathop{\sum_{u_1,\ldots,u_N\in U}}_{v_1,\ldots,v_N\in V}
\left|\det\left(\frac{1}{u_i-v_j}\right)_{(i,j)\in\{1,\ldots,N\}^2}
\right|^2
\eq
where the sums are defined with the prescription that configurations
which differ only by a permutation of identical particles are counted
only once, and $\lambda_p$ is the constant fugacity of the charges of
species $p=\pm$. This grand partition function is the determinant of
an anti Hermitian matrix $\mathbf{M}$ explicitly shown in
Ref. \cite{Cornu89}. 

When passing to the continuum limit in the element $\mathbf{M}_{ij}$
one should replace $u_i$ or $v_i$ by $z$ and $u_j$ or $v_j$ by
$z^\prime$, {\sl i.e.} $i\to z$ and $j\to z^\prime$. 
Each lattice site is characterized by its complex coordinate $z$ and
an isospinor which is \mbox{\tiny
  $\left(\begin{array}{c}1\\0\end{array}\right)$}  
if the site belongs to the positive sub-lattice $U$ and
\mbox{\tiny $\left(\begin{array}{c}0\\1\end{array}\right)$} if it
belongs to the negative sub-lattice $V$. We then define a matrix $\pmb{\mathcal
  M}$ by
\bq \label{M}
\langle z|\pmb{\mathcal M}|z^\prime\rangle=
\frac{\pmb{\sigma_x}+i\pmb{\sigma_y}}{2}\frac{2L/M}{z-z^\prime}+
\frac{\pmb{\sigma_x}-i\pmb{\sigma_y}}{2}
\frac{2L/M}{\overline{z}-\overline{z}^\prime}~,
\eq
where the $\pmb{\sigma}$'s are the $2\times 2$ Pauli matrices operating in
the isospinor space, and $z=x+iy$.

The matrix $\pmb{\mathcal M}$ can be expressed in terms of a simple Dirac operator
\bq
\langle z|\pmb{\mathcal M}|z^\prime\rangle=\frac{2L}{M}
(\pmb{\sigma_x}\partial_x+\pmb{\sigma_y}\partial_y)
\ln|z-z^\prime|~,
\eq
and the grand partition function can be rewritten as 
\bq \nonumber
\Xi&=&\det\left\{\mathbf{1}\delta^{(2)}(z;z^\prime)
+\left[\lambda_+\frac{\mathbf{1}+\pmb{\sigma_z}}{2}+
\lambda_-\frac{\mathbf{1}-\pmb{\sigma_z}}{2}\right]
\langle z|\pmb{\mathcal M}|z^\prime\rangle\right\}\\ \label{gpf}
&=&\det[\mathbf{I}+\mathbf{K}^{-1}]~,
\eq
with $\mathbf{I}=\mathbf{1}\delta^{(2)}(z;z^\prime)$ and
\bq
\pmb{\lambda}&=&\lambda_+\frac{\mathbf{1}+\pmb{\sigma_z}}{2}+
\lambda_-\frac{\mathbf{1}-\pmb{\sigma_z}}{2}~,\\
\mathbf{K}^{-1}&=&\pmb{\lambda}\pmb{\mathcal M}~.\\
\eq

Then, since
$\Delta_{\mbox{flat}}\ln|z|=2\pi\delta(s)\delta(\varphi)/s=2\pi\delta(z)$
(where $s=|z|$ and $\delta(z)$ is the flat Dirac delta function), the
inverse operator $\mathbf{K}$ is  
\bq
\mathbf{K}=\mathbf{m}^{-1}(z)\mathbf{O}~,
\eq 
where
\bq 
\mathbf{m}(z)&=&m_+(z)\frac{\mathbf{1}+\pmb{\sigma_z}}{2}+
m_-(z)\frac{\mathbf{1}-\pmb{\sigma_z}}{2}~,\\
\mathbf{O}&=&\frac{2}{M}(\pmb{\sigma_x}\partial_x+
\pmb{\sigma_y}\partial_y)~.
\eq
Here $m_{\pm}(z)=(2\pi L\lambda_\pm\sqrt g/sS)(2/M)^2$ are rescaled
position dependent fugacities, $g=\det(g_{\mu\nu})$, and $S$ is the
area per lattice site which appears when the discrete sums are
replaced by integrals. 
Notice that in the flat limit $M\to 0$ we 
find $m_{\pm}\to \widetilde{m}_\pm$, where $\widetilde{m}_\pm=2\pi
L\lambda_\pm/S$ are the rescaled fugacities of the flat system
\cite{Cornu89,Cornu87}. Moreover $S$ is a local property of the
surface independent from its curvature.

We then find 
\bq \nonumber
\ln\Xi&=&\mbox{Tr}\left\{\ln\left[\mathbf{I}+\mathbf{K}^{-1}
\right]\right\}~,
\eq
and the one-body densities and $n$-body truncated densities can be
obtained in the usual way by taking functional derivatives of the
logarithm of the grand partition function with respect to the
fugacities $\lambda_{\pm}$. Marking the sign of the particle charge at
$z_i$ by an index $p_i=\pm1$, and defining the matrix 
\bq \label{Green}
R_{p_1p_2}(z_1,z_2)=\langle z_1p_1|
\mathbf{K}^{-1}(\mathbf{I}+\mathbf{K}^{-1})^{-1}
|z_2p_2\rangle~,
\eq
it can then be shown
\cite{Cornu87,Cornu89} that they are given by
\bq \label{rho1}
&\rho_{p_1}^{(1)}(z_1)=
R_{p_1p_1}(z_1,z_1)~,&\\ \label{rho2}
&\rho_{p_1p_2}^{(2)T}(z_1,z_2)=
-R_{p_1p_2}(z_1,z_2)R_{p_2p_1}(z_2,z_1)~,&\\ \nonumber
&\rho_{p_1p_2,\ldots,p_n}^{(n)T}
(z_1,z_2,\ldots,z_n)=&\\
&(-)^{n+1}\sum_{(i_1,i_2,\ldots,i_n)}
R_{p_{i_1}p_{i_2}}(z_{i_1},z_{i_2})\cdots
R_{p_{i_n}p_{i_1}}(z_{i_n},z_{i_1})~,&
\eq
where $m_p(z)$ $=$ $(2\pi L\lambda_p\sqrt g/sS)(2/M)^2$ $=$
$(\widetilde{m}_p\sqrt g/s)(2/M)^2$, $\widetilde{m}_p$ $=$ $2\pi L\lambda_p/S$,
$\sqrt g$ $=$ $(M/2)^2s(1+1/s)^4$, and the summation runs over
all cycles $(i_1,i_2,\ldots,i_n)$ built with $\{1,2,\ldots,n\}$. 

\subsection{Symmetries of the Green's function $R$}

Since $\mathbf{m}^\dagger (z) =\mathbf{m}(z)$ and $\mathbf{O}^\dagger
=-\mathbf{O}$ we find 
\bq
\overline{R_{p_1p_2}(z_1,z_2)}=\langle z_2p_2|\mathbf{m}^{-1}\mathbf{O}
(\mathbf{I}-\mathbf{m}^{-1}\mathbf{O})^{-1}(\mathbf{m}^{-1}\mathbf{O})^{-1}
|z_1p_1\rangle~.
\eq
Expanding in $\mathbf{O}$ (by using 
$(\mathbf{I}-\mathbf{m}^{-1}\mathbf{O})^{-1}=\mathbf{I}+
\mathbf{m}^{-1}\mathbf{O}+(\mathbf{m}^{-1}\mathbf{O})^2+\ldots$)  
and comparing with the definition
$R_{p_1p_2}(z_1,z_2)=\langle z_1p_1|(\mathbf{m}^{-1}\mathbf{O})^{-1}
(\mathbf{I}+\mathbf{m}^{-1}\mathbf{O})^{-1}\mathbf{m}^{-1}
\mathbf{O}|z_2p_2\rangle$ we find
\bq
\overline{R_{pp}(z_1,z_2)}=R_{pp}(z_2,z_1)~,\\
\overline{R_{p-p}(z_1,z_2)}=-R_{-pp}(z_2,z_1)~.
\eq
From which also follows that $R_{pp}(z_1,z_1)$ has to be real. If
$\lambda_+=\lambda_-$ then we additionally must have
\bq
R_{pp}(z_1,z_2)=R_{-p-p}(z_1,z_2)~.
\eq

\subsection{Two-body truncated correlation functions and perfect
  screening sum rule}

For the two-body truncated correlation functions of Eq. (\ref{rho2})
we then find
\bq \label{rhopp}
\rho_{++}^{(2)T}(z_1,z_2)&=&-|R_{++}(z_1,z_2)|^2~,\\ \label{rhopm}
\rho_{+-}^{(2)T}(z_1,z_2)&=&|R_{+-}(z_1,z_2)|^2~.
\eq

Notice that the total correlation function for the like particles
$h_{++}(z_1,z_2)$ $=$ $\rho_{++}^{(2)T}(z_1,z_2)$ $/
\rho_{+}^{(1)}(z_1)\rho_{+}^{(1)}(z_2)$ goes to $-1$ when the
particles coincide $z_1\to z_2$ as follows from the structure of
Eqs. (\ref{rho1})-(\ref{rho2}). Moreover the truncated densities of 
any order has to decay to zero as two groups of particles are
infinitely separated. In particular
$|R_{++}(z_1,z_2)|=|R_{++}(s1,s2;\varphi)|$, with
$\varphi=\varphi_1-\varphi_2$,   
has to decay to zero as $|s_1-s_2|\to \infty$. 

The perfect screening sum rule has to be satisfied for the symmetric
mixture 
\bq \label{perfect-screening}
\int[\rho_{+-}^{(2)T}(z_1,z_2)-\rho^{(2)T}_{++}(z_1,z_2)]\sqrt{g_1}
ds_1d\varphi_1=\rho_\pm(z_2)~,
\eq
where $g_1$ is $g$ calculated on particle $1$.

\subsection{From the structure to the thermodynamics}

Following Ref. \cite{Tellez98}, we note that if we solve the
eigenvalue problem 
$\mathbf{K}\pmb{\psi}_n=\lambda_n\pmb{\psi}_n$ (here
$\pmb{\psi}_n$ is a spinor of components $\psi_{+,n}$ and $\psi_{-,n}$),
then 
\bq \label{gpfT}
\Xi=\prod_n\left(1+\frac{1}{\lambda_n}\right)~.
\eq

Now the eigenvalue problem reads 
\bq \label{psi-}
\left\{m_+(z)+A [m_-(z)]^{-1}A^\dagger
\right\}\psi_-(z)=(m_-(z)-\lambda^2)\psi_-(z)~,\\ \label{psi+}
\left\{m_-(z)+A^\dagger [m_+(z)]^{-1}A
\right\}\psi_+(z)=(m_-(z)-\lambda^2)\psi_+(z)~,
\eq
where $A=(2/M)(\partial_{x}+i\partial_{y})$ and
$A^\dagger=(2/M)(-\partial_{x}+i\partial_{y})$ and the left hand
side is the same as the left hand side for the Green's functions
Eqs. (\ref{G++})-(\ref{G--}).

Of course we know that the thermodynamic quantities all diverge in our
$\Gamma=2$ case so Eq. (\ref{gpfT}), although suggestive, is of small
practical interest.

\subsection{Determination of the Green's function $R$}

The Green's function matrix $\mathbf{R}$ is the solution of a system of
four coupled partial differential equations, namely
\bq
(\mathbf{I}+\mathbf{K}^{-1})\mathbf{K} \mathbf{R}(z_1,z_2)=
(\mathbf{I}+\mathbf{K}) \mathbf{R}(z_1,z_2)=
\mathbf{1}\delta^{(2)}(z_1;z_2)
\eq
where $\delta^{(2)}(z_1;z_2)$ $=$
$\delta^{(2)}(s,\varphi;s_0,\varphi_0)$ $=$ $(\sqrt g)^{-1}
\delta(s-s_0)\delta(\varphi-\varphi_0)$ is the Dirac delta function on
the curved surface, $\delta(s-s_0)\delta(\varphi-\varphi_0)/s$ $=$
$\delta(s,\varphi;s_0,\varphi_0)$ $=$ $\delta(z;z_0)$ is the flat
Dirac delta function, and $\mathbf{1}$ is the identity matrix. These can
be rewritten as follows
\bq \nonumber
\left[\mathbf{O}+\mathbf{m}(z_1)\right]
\mathbf{R}(z_1,z_2)=\mathbf{m}(z_1)\delta^{(2)}(z_1;z_2)~.
\eq

If instead of $\mathbf{R}$ one uses
$\mathbf{R}=\mathbf{G}\mathbf{\widetilde{m}}$, $\mathbf{G}$ satisfies the
equation
\bq \label{Geq}
[\mathbf{O}+\mathbf{m}(z_1)]\mathbf{G}(z_1,z_2) =
\frac{4}{M^2}\mathbf{1}\delta(z_1;z_2)~,  
\eq 
where $\delta(z_1;z_2)$ is now the flat Dirac delta function
\cite{gtroot}. 

By combining the components of this equation one obtains decoupled
equations for $G_{++}$ and $G_{--}$ as follows
\bq \label{G++}
\left\{m_+(z_1)+A^\dagger [m_-(z_1)]^{-1}A
\right\}G_{++}(z_1,z_2)&=&
\frac{4}{M^2}\delta(s_1,\varphi_1;s_2,\varphi_2)~,\\ \label{G--}
\left\{m_-(z_1)+A[m_+(z_1)]^{-1}A^\dagger
\right\}G_{--}(z_1,z_2)&=&
\frac{4}{M^2}\delta(s_1,\varphi_1;s_2,\varphi_2)~,
\eq
while
\bq \label{G-+}
G_{-+}(z_1,z_2)&=&-\left[m_-(z_1)\right]^{-1}
AG_{++}(z_1,z_2)~,\\
G_{+-}(z_1,z_2)&=&+\left[m_+(z_1)\right]^{-1}
A^\dagger G_{--}(z_1,z_2)~,
\eq

Then Eq. (\ref{G++}) can be rewritten in Cartesian coordinates as
\bq \nonumber
&&\left\{m_+(z_1)m_-(z_1)-\left(\frac{2}{M}\right)^2
\left[(\partial_{x_1}^2+\partial_{y_1}^2)-\right.\right.\\ \nonumber
&&\left.\left.\frac{4(-x_1+iy_1)}{s_1^2(1+s_1)}
(\partial_{x_1}+i\partial_{y_1})\right]\right\}G_{++}(z_1,z_2)= 
\\ \nonumber
&&\left(\frac{2}{M}\right)^4
\frac{\widetilde{m}_-\sqrt{g_1}}{s_1^2}\delta(s_1-s_2)
\delta(\varphi_1-\varphi_2)=\\ \label{G++cart}
&&\left(\frac{2}{M}\right)^4\frac{\widetilde{m}_-\sqrt{g_1}}
{\sqrt{x_1^2+y_1^2}}
\delta(x_1-x_2)\delta(y_1-y_2)~,
\eq
where $s=\sqrt{x^2+y^2}$. From the
expression of the gradient in polar coordinates follows  
\bq
\left\{\begin{array}{c} \displaystyle
\partial_x=\cos\varphi\partial_s
-\frac{\sin\varphi}{s}\partial_\varphi~,\\ \displaystyle
\partial_y=\sin\varphi\partial_s+\frac{\cos\varphi}{s}\partial_\varphi~.
\end{array}\right.
\eq
Which allows us to rewrite Eq. (\ref{G++cart}) in polar coordinates as  
\bq \nonumber
&&\left[\widetilde{m}_+\widetilde{m}_-\left(1+\frac{1}{s_1}\right)^8-
\left(\frac{2}{M}\right)^2\left(\frac{1}{s_1}\partial_{s_1}(s_1\partial_{s_1})+
\frac{1}{s_1^2}\partial^2_{\varphi_1}+\right.\right. \\ \nonumber
&&\left.\left.\frac{4}{s_1(1+s_1)}\partial_{s_1}+
\frac{4i}{s_1^2(1+s_1)}\partial_{\varphi_1}\right)\right] 
G_{++}(z_1,z_2)=\\ \label{G++polar}
&&\left(\frac{2}{M}\right)^4\frac{\widetilde{m}_-\sqrt{g_1}}{s_1^2}
\delta(s_1-s_2)\delta(\varphi_1-\varphi_2)~.
\eq
From this equation we immediately see that $G_{++}(z_1,z_2)$ cannot be real.
Notice that in the flat limit $M\to 0$ we have $s\sim 2r/M$ and
Eq. (\ref{G++polar}) reduces to  
\bq \nonumber
&&\left[\widetilde{m}_+\widetilde{m}_--
\frac{1}{r_1}\partial_{r_1}(r_1\partial_{r_1})-
\frac{1}{r_1^2}\partial^2_{\varphi_1}\right]G_{++}(z_1,z_2)=
\\ \label{G++flateq}
&&\frac{\widetilde{m}_-}{r_1}\delta(r_1-r_2)\delta(\varphi_1-\varphi_2)~.
\eq
which, when $\widetilde{m}_+=\widetilde{m}_-=\widetilde{m}$, has the following
well known solution \cite{Cornu89,Cornu87} 
\bq
G_{++}(z_1,z_2)=\frac{\widetilde{m}}{2\pi}K_0(\widetilde{m}|\rr_1-\rr_2|)~,
\eq
where $K_0$ is a modified Bessel function.

Let us from now on restrict to the case of equal fugacities of the two
species. Then $\lambda_-=\lambda_+=\lambda$ with  
\bq
\widetilde{m}=\frac{2\pi L}{S}\lambda
=\frac{2\pi Le^{\beta\mu}}{\Lambda^2}
=\left(2\pi L\frac{mq^2}{4\pi\hbar^2}\right)e^{2\mu/q^2}~,
\eq
where $\hbar$ is Planck's constant, $m$ is the mass of the particles,
and $\mu$ the chemical potential. So $\widetilde{m}$ has the dimensions of
an inverse length. From the symmetry of the problem we can say that
$G_{++}=G_{++}(s_1,s_2;\varphi_1-\varphi_2)$. We can then express the
Green's function as the following Fourier series expansion 
\bq \label{fourier-expansion}
G_{++}(s_1,s_2;\varphi)=\frac{1}{2\pi}
\sum_{k=-\infty}^\infty g_{++}(s_1,s_2;k)e^{ik\varphi}~.
\eq 
Then, using the expansion of the Dirac delta function, $\sum_k
e^{ik\varphi}=2\pi\delta(\varphi)$, we find that $g_{++}$, a
continuous real function symmetric under exchange of $s_1$ and $s_2$,
has to satisfy the following equation    
\bq \nonumber
\left[Q_0\left(k,s_1\right)
+Q_1\left(s_1\right) \partial_{s_1}
+ Q_2\left(s_1\right) \partial_{s_1}^2\right]
g_{++}\left(s_1,s_2;k\right)=\\ \label{Eq0} 
\left(\frac{2}{M}\right)^2\widetilde{m}s_1^3(1+s_1)^5
\delta\left(s_1-s_2\right)~,  
\eq
where
\bq \nonumber
Q_0\left(k,s\right)&=&{\widetilde{m}}^2 \left(1+s\right)^9+
\left(\frac{2}{M}\right)^2
ks^6\left(4+k\left(1+s\right)\right)~,\\ \nonumber 
Q_1\left(s \right)&=&-\left(\frac{2}{M}\right)^2s^7\left(5+s\right)~,
\\ \nonumber
Q_2\left(s \right)&=&-\left(\frac{2}{M}\right)^2s^8\left(1+s\right)~.
\eq
And the coefficients $Q_i$ are polynomials of up to degree 9.
\subsection{Method of solution}

We start from the homogeneous form of Eq. (\ref{Eq0}). We note that,
for a given $k$, the two linearly independent solutions 
$f_\alpha(s;k)$ and $f_\beta(s;k)$ of this linear homogeneous second
order ordinary differential equation are not 
available in the mathematical literature to the best of our knowledge.
Assuming we knew those solutions we would then find the Green's function,
$g_{++}(s_1,s_2;k)$, writing \cite{Jackson}
\bq
f(t_1,t_2;k)=c_kf_\alpha(s_<;k)f_\beta(s_>;k)~,
\eq
where $s_<=\min(s_1,s_2)$, $s_>=\max(s_1,s_2)$, and $f_\beta$ has the
correct behavior at large $s$. Then
we determine $c_k$ by imposing the kink in $f$ due to the Dirac delta
function at $s_1=s_2$ as follows 
\bq
\partial_{s_1} f(s_1,s_2;k)|_{s_1=s_2+\epsilon}-\partial_{s_1}
f(s_1,s_2;k)|_{s_1=s_2-\epsilon}=-\widetilde{m}\frac{(1+s_2)^4}{s_2^5}~,
\eq 
where $\epsilon$ is small and positive.

The Green's function, symmetric under exchange of $s_1$ and $s_2$, is
reconstructed as follows  
\bq 
G_{++}(z_1,z_2)&=&G_{++}(s_1,s_2;\varphi)=
\frac{1}{2\pi}\sum_{k=-\infty}^\infty
c_kf_\alpha(s_<;k)f_\beta(s_>;k)e^{ik\varphi} \label{Gppms}
\eq
\section{TCP in the whole surface}
\label{sec:ws}

On the whole surface, using Eq. (\ref{vcoulombws}), we can now write
$e^{-\beta U_{2N}}$ at a coupling constant $\Gamma=2$ as follows
\bq
\left|\det\left(\frac{L}{L_0}
\frac{\sqrt{|u_j v_j|}}{u_i-v_j}\right)
_{(i,j)\in\{1,\ldots,N\}^2}\right|^2~.
\eq
The grand partition function will then be
\bq
\Xi=\det\left[\mathbf{I}+\mathbf{K_{\mbox{ws}}}^{-1}\right]~,
\eq
with 
\bq 
\mathbf{K_{\mbox{ws}}}^{-1} &=& \left(\frac{M}{2L_0}\right)
\pmb{\lambda}_{\mbox{ws}}\pmb{\mathcal{M}}~,\\ 
\pmb{\lambda}_{\mbox{ws}}&=&\lambda_+|z|
\frac{\mathbf{1}+\pmb{\sigma_z}}{2}+
\lambda_-|z|
\frac{\mathbf{1}-\pmb{\sigma_z}}{2}~.
\eq
The inverse operator is then given by
\bq 
\mathbf{K_{\mbox{ws}}}&=&\left(\frac{2L_0}{M}\right)
\pmb{\mathcal{M}}^{-1}\pmb{\lambda}_{\mbox{ws}}^{-1}~,\\
\pmb{\lambda}_{\mbox{ws}}^{-1}&=&\frac{1}{\lambda_+|z|}
\frac{\mathbf{1}+\pmb{\sigma_z}}{2}+
\frac{1}{\lambda_-|z|}
\frac{\mathbf{1}-\pmb{\sigma_z}}{2}~,\\
\pmb{\mathcal{M}}^{-1}&=&\frac{Ss}{2\pi L\sqrt{g}}
\left(\frac{M}{2}\right)^2\mathbf{O}~.
\eq
Introducing position dependent fugacities \cite{wsmethod} 
\bq
m_p(z)=\frac{2\pi (L/L_0)\lambda_p\sqrt g}{Ss}
=\widetilde{m}_p\frac{\sqrt{g}}{s}~,
\eq
where now $\widetilde{m}_p/L_0\to\widetilde{m}_p$, we can rewrite
\bq
\mathbf{K_{\mbox{ws}}}&=&
\frac{\pmb{\sigma_x}+i\pmb{\sigma_y}}{2}a_-+
\frac{\pmb{\sigma_x}-i\pmb{\sigma_y}}{2}a_+~,
\eq
with the operators
\bq
a_-&=&-\frac{\overline{z}}{m_-(z)|z|^3}+
\frac{1}{m_-(z)|z|}(\partial_x-i\partial_y)~,\\
a_+&=&-\frac{z}{m_+(z)|z|^3}+
\frac{1}{m_+(z)|z|}(\partial_x+i\partial_y)~.
\eq
Then the equation for the Green's functions are
\bq
(1-a_-a_+)R_{++}(z_1,z_2)=\delta^{(2)}(z_1;z_2)~,\\
(1-a_+a_-)R_{--}(z_1,z_2)=\delta^{(2)}(z_1;z_2)~,\\
R_{+-}=-a_-R_{--}~,\\
R_{-+}=-a_+R_{++}~.
\eq
The equation for $R_{++}$ in the symmetric mixture case is 
\bq \nonumber
\left[m^2(z_1)-\frac{2}{s_1^4}+\frac{2\partial_{s_1}}{s_1^3}
-\frac{\partial_{s_1}^2}{s_1^2}
-\frac{-i\partial_{\varphi_1}+\partial_{\varphi_1}^2}{s_1^4}
\right]R_{++}(z_1,z_2)=\\ \label{Eq1}
\frac{m^2(z_1)}{\sqrt{g_1}}
\delta(s_1-s_2)\delta(\varphi_1-\varphi_2)=
\frac{\widetilde{m}^2\sqrt{g_1}}{s_1^2}
\delta(s_1-s_2)\delta(\varphi_1-\varphi_2)~,
\eq
From this equation we see that $R_{++}(z_1,z_2)$ will now be real. 

By expanding Eq. (\ref{Eq1}) in a Fourier series in the azimuthal
angle we now find  
\bq \nonumber
\left[Q_0\left(k,s_1\right)+
Q_1(s_1)\partial_{s_1}+ Q_2\left(s_1\right)\partial_{s_1}^2\right]
g_{++}\left(s_1,s_2;k\right)=\\ \label{Eq2} 
\left(\frac{M}{2}\right)^2\widetilde{m}s_1^3(1+s_1)^4
\delta\left(s_1-s_2\right)~,  
\eq
where
\bq \nonumber
Q_0\left(k,s\right)&=&\left(\frac{M}{2}\right)^4{\widetilde{m}}^2 
\left(1+s\right)^8+s^4(k^2-k-2)~,\\ \nonumber
Q_1(s)&=&2s^5~,\\ \nonumber
Q_2\left(s \right)&=&-s^6~.
\eq
And the coefficients $Q_i$ are now polynomials of up to degree 8.

In the flat limit we find, for $G_{++}=R_{++}/\widetilde{m}$,
the following equation 
\bq \nonumber
\left[\widetilde{m}^2-\frac{2}{r_1^4}+\frac{2\partial_{r_1}}{r_1^3}-
\frac{\partial^2_{r_1}}{r_1^2}-
\frac{-i\partial_{\varphi_1}+\partial^2_{\varphi_1}}{r_1^4}
\right]G_{++}(z_1,z_2)=\\
\frac{\widetilde{m}}{r_1}\delta(r_1-r_2)\delta(\varphi_1-\varphi_2)~.
\eq
We then see that we now do not recover the TCP in the
plane \cite{Cornu89,Cornu87}. This has to be expected because in the
flat limit the Flamm's paraboloid reduces to two planes connected by
the origin. 

After the Fourier expansion of Eq. (\ref{fourier-expansion}) we now
get  
\bq \label{Eq3}
[P_0(k,r_1)+P_1(r_1)\partial_{r_1}+P_2(r_1)\partial_{r_1}^2]
g_{++}(r_1,r_2;k)=\widetilde{m}\delta(r_1-r_2)~,
\eq
where
\bq \nonumber
P_0(k,r)&=&\widetilde{m}^2r+\frac{k^2-k-2}{r^3}~,\\ \nonumber
P_1(r)&=&\frac{2}{r^2}~,\\ \nonumber
P_2(r)&=&-\frac{1}{r}~.
\eq
The homogeneous form of this equation admits the following two
linearly independent solutions 
\bq \nonumber
&&\left.\begin{array}{ll}
f_1(r;-1)=&[D_{-1/2}(i\sqrt{2\widetilde{m}}r)
+\overline{D_{-1/2}(i\sqrt{2\widetilde{m}}r)}]/2\\
f_2(r;-1)=&D_{-1/2}(\sqrt{2\widetilde{m}}r)\\
\end{array}\right\}~~~k=-1~,\\ \nonumber
&&\left.\begin{array}{ll}
f_1(r;2)=&[D_{-1/2}((-2)^{1/4}\sqrt{\widetilde{m}}r)+\\
&\overline{D_{-1/2}((-2)^{1/4}\sqrt{\widetilde{m}}r)}]/2\\
f_2(r;2)=&[D_{-1/2}(i(-2)^{1/4}\sqrt{\widetilde{m}}r)+\\
&\overline{D_{-1/2}(i(-2)^{1/4}\sqrt{\widetilde{m}}r)}]/2
\end{array}\right\}~~~k=2~,\\ \nonumber
&&\left.\begin{array}{ll}
f_1(t;k)=&\sqrt rI_{-\sqrt{7-4k+4k^2}/4}(\widetilde{m}r^2/2)\\
f_2(t;k)=&\sqrt rI_{\sqrt{7-4k+4k^2}/4}(\widetilde{m}r^2/2)
\end{array}\right\}~~~\mbox{else}~,
\eq 
where $D_\nu(x)$ are parabolic cylinder functions and $I_\mu(x)$ are the
modified Bessel functions of the first kind which diverge as
$e^x/\sqrt{2\pi x}$ for large $x\gg|\mu^2-1/4|$. 

Again we write $g_{++}(r_1,r_2;k)=c_kf_\alpha(r_<;k)f_\beta(r_>;k)$
and impose the kink condition,
\bq
\partial_{r_1}g_{++}(r_1,r_2;k)|_{r_1=r_2+\epsilon}-
\partial_{r_1}g_{++}(r_1,r_2;k)|_{r_1=r_2-\epsilon}=
-\widetilde{m}r_2~,
\eq 
to find the $c_k$. The Green's function is then reconstructed using
Eq. (\ref{Gppms}). But we immediately see that curiously 
$|G_{++}|$ diverges. Even
the structure of the plasma is not well defined in this situation. The
collapse of opposite charges at the horizon shrinking to the origin
makes the structure of the plasma physically meaningless. 

\section{Conclusions}
\label{sec:conclusions}

An attempt was made to extend the solution of the one component plasma,
at the special value of the coupling constant $\Gamma=2$, living in the
Flamm's paraboloid \cite{Fantoni2008} to the two component one, the TCP.
The Flamm's paraboloid is a Riemannian surface with non-constant
curvature which is asymptotically flat. Its curvature depends on a
parameter $M$ ($2M$ being the radius of the horizon) in such way that
for $M\to 0$ the surface becomes flat: two planes connected by the
origin. 

The work of F. Cornu and B. Jancovici \cite{Cornu89,Cornu87} on the
TCP in a plane showed that at such particular value of 
the coupling constant, using a Cauchy's identity, it is possible to
find an analytic solution to the structure of the plasma starting from
a discretized model to avoid divergences and taking the continuous
limit in the end. All the $n$-particle correlation functions with
$n>1$ are well behaved except the particle density which diverges as
all the other thermodynamic quantities due to the competition between
the attraction of oppositely charged particles and the thermal
motion. This can be prevented by adding an hard core to the particles
in order to prevent the collapse.

In this work we repeated the same calculation for the plasma living
in the Flamm's paraboloid, in the half surface case, $s>1$, with an
insulating horizon (at $s=1$) and in the full surface case,
$s>0$. The solution of the equation determining the structure, the
Green's function equation, is reduced to the mathematical problem of
finding the 
two linearly independent solutions of a linear homogeneous second
order ordinary  differential equation with polynomial coefficients of
high degree (9th in the half surface case and 8th in the full surface
case). To the best of our knowledge, an analytic solution of the
equations turns out to be out of mathematical reach. Nonetheless the
many-body problem has been reduced to finding the solutions of a
simple differential equation. The further step of finding explicitly
the alanytical form of the solutions from the given differential
equations is a matter of mathematical syntax and we think does not add
much to the present work.   

We discussed the symmetries of the Green's function, the perfect
screening sum rule, the relationship between the structure and
the thermodynamics, and the method of solution of the Green's function
equation. 

We found that for the plasma living in the half Flamm's paraboloid with
an insulating horizon the flat limit $M\to 0$ reduces the system to
the TCP on the plane \cite{Cornu89,Cornu87}.
For the plasma living in the full Flamm's paraboloid 
taking the flat limit one does not recover the plasma in the plane and
this has to be expected since the paraboloid in this limit reduces to
two planes connected by a hole at the origin.
Instead the resulting solution for the structure turns out to be
physically meaningless. The collapse of opposite charges at the
horizon as it shrinks to the origin spoils the structure of the
plasma. 

We leave as an open problem the one of finding approximations to the
two linearly independent solutions of the homogeneous counterpart of
Eq. (\ref{Eq0}), for the half surface case, and of Eq. (\ref{Eq2}),
for the whole surface case, as a function of the integer $k$, the 
further determination of the Green function in these two cases,
through Eq. (\ref{Gppms}), the eventual analysis of the solution and
discussion about the necessity of the addition of a hard core
to the particles (hard ``disks'' of diameter $D$ with $\delta
s<2D(1+1/s)^{-2}/M$) in order to cure the divergence of the
densities $\rho_p$ with the consequent determination of the equation
of state. These final steps are necessary in order to have the 
problem of the structure of the fluid reduced to a closed 
form. In the present work we limited ourselves to what can be said in
an analytical form.



\ack
I would like to acknowledge the financial support and hospitality from
the National Institute for Theoretical Physics of South Africa during the
preparation of the work. I would also like to thank Peter Forrester
for stimulating the work and Filippo Giraldi for discussions on the
solution of the Green's function equation.  
\section*{References}

\begin{thebibliography}{10}
\bibitem{gtroot}
The same result could have been reached by noticing that the partition function
  of Eq. (\ref{pf0}) rewritten in the continuum is equivalent to the partition
  function of the flat system with position dependent fugacities
  $\zeta_p(s)=\lambda_p(M/2)^2(1+1/s)^4(2L/M)$. This leads naturally
  \cite{Cornu89} to define a green function
  $\mathbf{\tilde{G}}=(M/2)\mathbf{G}$ which satisfies the equation
  $[(M/2)\mathbf{O}+(2\pi/S)\pmb{\zeta}(z)]\mathbf{\tilde{G}}(z,z^\prime)=\mathbf{1}\delta(z;z^\prime)$
  which reduces to Eq. (\ref{Geq}).
\bibitem{wsmethod}
Alternatively we could have left the one body terms ($\sqrt{|z|}$) outside the
  determinant, and then consider them as part of the position depend
  fugacities, and simply apply F. Cornu and B. Jancovici \cite{Cornu89}
  formalism with these position depend fugacities.
\end{thebibliography}

\end{document}